\documentclass[12pt]{article}
\usepackage{amssymb}
\usepackage{latexsym}
\usepackage{listings}
\usepackage{xcolor}
\usepackage{placeins}
\usepackage{natbib}
\usepackage{url}
\usepackage{multirow}
\usepackage{dblfloatfix}
\usepackage{subcaption}
\usepackage{tikz}
\usetikzlibrary{shapes.geometric, arrows, positioning, shadows.blur}
\usepackage{lineno}
\tikzstyle{block} = [rectangle, rounded corners, draw, fill=blue!20, blur shadow={shadow blur steps=5}, text width=3cm, text centered, minimum height=1cm]
\tikzstyle{line} = [draw, -latex', thick]
\definecolor{backgroundColour}{rgb}{0.95,0.95,0.92}
\definecolor{commentColour}{rgb}{0.25,0.50,0.37}
\definecolor{stringColour}{rgb}{0.16,0.00,1.00}
\definecolor{keywordColour}{rgb}{0.37,0.08,0.25}
\definecolor{vimGreen}{rgb}{0,0.6,0}
\definecolor{mygreen}{rgb}{0,0.6,0}
\definecolor{mygray}{rgb}{0.5,0.5,0.5}
\definecolor{mymauve}{rgb}{0.58,0,0.82}
\lstdefinestyle{mystyle}{
    backgroundcolor=\color{backgroundColour},
    commentstyle=\color{commentColour},
    keywordstyle=\color{keywordColour},
    numberstyle=\tiny\color{gray},
    stringstyle=\color{stringColour},
    basicstyle=\footnotesize\ttfamily,
    breakatwhitespace=false,
    breaklines=true,
    captionpos=b,
    keepspaces=true,
    numbers=left,
    numbersep=5pt,
    showspaces=false,
    showstringspaces=false,
    showtabs=false,
    tabsize=2,
    language=Python
}
\begin{document}
\title{DNA Sequence Classification with Compressors}
\author{Şükrü Ozan\\
digiMOST GmbH, Dieselstraße  7, Marl, 45770, \\
Nordrhein-Westfalen, Deutschland\\
Email: \texttt{sukruozan@digimost.de}}
\maketitle
\begin{abstract}
Recent studies in DNA sequence classification have leveraged sophisticated machine learning techniques, achieving notable accuracy in categorizing complex genomic data. Among these, methods such as k-mer counting have proven effective in distinguishing sequences from varied species like chimpanzees, dogs, and humans, becoming a staple in contemporary genomic research. However, these approaches often demand extensive computational resources, posing a challenge in terms of scalability and efficiency. Addressing this issue, our study introduces a novel adaptation of Jiang et al.'s compressor-based, parameter-free classification method, specifically tailored for DNA sequence analysis. This innovative approach utilizes a variety of compression algorithms, such as Gzip, Brotli, and LZMA, to efficiently process and classify genomic sequences. Not only does this method align with the current state-of-the-art in terms of accuracy, but it also offers a more resource-efficient alternative to traditional machine learning methods. Our comprehensive evaluation demonstrates the proposed method's effectiveness in accurately classifying DNA sequences from multiple species. We present a detailed analysis of the performance of each algorithm used, highlighting the strengths and limitations of our approach in various genomic contexts. Furthermore, we discuss the broader implications of our findings for bioinformatics, particularly in genomic data processing and analysis. The results of our study pave the way for more efficient and scalable DNA sequence classification methods, offering significant potential for advancements in genomic research and applications.\end{abstract}


\section{Introduction} \label{S:1}
The pursuit of efficient deoxyribonucleic acid (DNA) sequence classification is an important operation in genomic research, and as a natural consequence of recent advances in artificial intelligence, it has been significantly driven by machine learning. Traditional methods, more specifically k-mer counting and thresholding, have been widely used to effectively classify DNA sequences.

Traditional methods of processing DNA sequences often result in vectors of varying lengths, which poses a challenge for classification or regression algorithms that require uniform length inputs. To overcome this, sequences are typically truncated or padded. DNA and protein sequences can be conceptualized as a language, with the genome as a book, subsequences as sentences and chapters, k-mers as words, and nucleotide bases as alphabets. This analogy to natural language suggests the potential applicability of natural language processing (NLP) methods to genomic sequences.

The k-mer counting method involves decomposing long biological sequences into overlapping  sequences (`words') of length `k'. For example, a sequence like ``ATGCATGCA'' is broken down into k-mers of length 6 (hexamers) to produce ``ATGCAT'', ``TGCATG'', GCATGC'', ``CATGCA'', where `A', `G', `C' and `T' represent nitrogenous bases found in DNA. Python's natural language processing tools can facilitate the k-mer counting process, making it manageable and straightforward. There are numerous works publicly shared by machine learning enthusiasts on DNA sequence classification using k-mer counting, such as \citet{Nageshsingh:2023:1}.

Juneja et al.'s exploration of k-mer counting in DNA sequence classification \citet{Juneja:2022} highlights its efficacy in distinguishing species-specific sequences, emphasizing the influence of k-mer size on classification accuracy. This method was used in \citet{OrozcoArias:2021} to classify long terminal repeats retrotransposons (LTRs) from genomic sequences obtained from a public plant genome database.

In a parallel line of research focusing on feature selection in microRNA data, a significant contribution has been made by employing advanced meta-heuristic algorithms. A noteworthy study in this realm is presented by \citet{Jaddi:2022}, where the authors enhanced the cell separation meta-heuristic algorithm (CSA) for effective feature selection in cancer classification. This improved approach, termed I-CSA, demonstrated remarkable performance in selecting discriminative features from high-dimensional genomic data, achieving an outstanding classification accuracy. Their methodology underscores the importance of efficient feature selection in genomic research, providing valuable insights that complement traditional methods.

In \citet{Sarkar:2021}, the authors employed the k-mer method in a novel way for information recovery from DNA sequences. They developed a three-part algorithm that uses a finite impulse response digital filter to calculate the density of k-mer or q-gram words in a sequence. This calculated density is then analyzed using principal component analysis to assess the dissimilarity between sequences, leading to the formation of clusters that aid in constructing phylogenetic relationships.

A recent study by Wen et al. \citet{Wen:2019} integrated k-mer counting with convolutional neural networks to enhance feature extraction in Long-chain non-coding ribonucleic acid (lncRNA) and messenger RNA (mRNA) classification. Furthermore, Arias et al. \citet{MillnArias:2022} demonstrated the potential of unsupervised learning techniques, such as clustering algorithms, in identifying patterns in genomic sequences.

The field of explainable AI (XAI) in bioinformatics, as discussed in Yagin et al.'s work \citet{Yagin:2023}, provides frameworks for interpreting the results of machine learning models on COVID-19 metagenomic next-generation sequencing (mNGS) samples. Additionally, researchers like Salman \citet{Khan:2020} are developing scalable solutions like cloud-based tools and parallel computing strategies to handle the increasing volume of genomic data.

In related literature, a study by Jiang et al. \citet{Jiang:2023} claims to achieve text classification accuracy comparable to deep neural networks (DNNs) using a solution that embodies Occam's Razor principle. This alternative method combines a compressor algorithm, like gzip, with a k-nearest-neighbor classifier, requiring no training parameters. This approach outperforms BERT on benchmark datasets and excels in few-shot settings where labeled data are too scarce to train DNNs effectively.

This study introduces a novel adaptation of Jiang et al.'s method tailored for DNA sequence classification. We explore the performance of various compressor algorithms, including Gzip, Snappy, Brotli, LZ4, Zstandard, BZ2, and LZMA, each with its unique strengths and suitability for DNA sequencing. This approach diverges from traditional deep learning models, offering a resource-efficient alternative that competes in accuracy.

We demonstrate the efficacy of these compressor algorithms in accurately classifying DNA sequences, proposing a significant advancement in genomic research. This method aligns with current state-of-the-art accuracy and addresses possible computational resource constraints in bioinformatics.

\section{Material and Methods}

Our methodology integrates the compressor-based classification method proposed by Jiang et al. with various compression algorithms for DNA sequence classification. The method employs a non-parametric approach using compressors and a k-nearest-neighbor classifier. This absence of training parameters makes it significantly resource-efficient. The method involves comparing a given sequence with all previously labeled (training) data to determine its similarity to each sample. The similarity metric is calculated by capturing the amount of compression rate after concatenating the given sequence with the samples in the training data. Since compression algorithms natively capture the similarities in a given sequence, the amount of compression constitutes a good metric to detect similarity between two sequences. 

In this study, the method is adapted for DNA sequences, where the sequence data is processed through different compressors like Gzip, Snappy, Brotli, LZ4, Zstandard, BZ2, and LZMA. Each compressor's performance is evaluated based on its specific algorithmic characteristics, such as compression speed, efficiency, and suitability for DNA sequence data from different species (i.e. human, chimpanzee and dog).

This approach allows us to ascertain the most effective compressor for DNA sequence classification, considering factors like computational resource constraints and the nature of genomic data. The detailed performance analysis of each compressor will provide insights into their applicability and efficiency in DNA sequence classification, contributing to the field of bioinformatics with a novel, resource-efficient classification method.

In the next section, we will further elaborate on the competitiveness of selected compressor algorithms by demonstrating classification accuracy assessments.

\subsection{Dataset}

The dataset used in this study is a collection of DNA sequences from a limited variety of organisms, namely human, chimpanzee and dog, sourced from a publicly available Kaggle dataset \citet{Nageshsingh:2023:2}. Each sequence in the dataset represents a segment of genetic material, encoded in a standard format with characters representing nucleotide bases namely Adenine (A), Thymine (T), Cytosine (C), and Guanine (G).

The sequences have been categorized to facilitate supervised learning tasks. Each sequence is labeled, allowing for clear classification and analysis. The sequences are presented in a format conducive to machine learning applications, with each nucleotide base represented by its corresponding character. This format simplifies the process of numerical or categorical encoding necessary for machine learning algorithms.

The categorization includes seven sequence classes, described as follows:

\begin{enumerate}
    \item \textbf{G protein-coupled receptors (GPCRs):} These are cell membrane proteins that mediate cellular responses to various signals like hormones and neurotransmitters. These protein structures are labelled as Class \#0 in the dataset.
    \item \textbf{Receptor tyrosine kinase (RTKs):} Involved in the cellular response to growth factors such as insulin. They are labelled as Class \#1 in the dataset.
    \item \textbf{Protein tyrosine phosphatases (PTPs):} These enzymes remove phosphate groups from tyrosine residues on proteins, counteracting the action of tyrosine kinases, and thus serve as regulatory elements in signaling pathways. These  structures are labelled as Class \#2 in the dataset. 
    \item \textbf{Synthetases:} Enzymes that catalyze the joining of two molecules, typically using ATP, often involved in the synthesis of larger biomolecules from smaller components. These  structures are labelled as Class \#3 in the dataset.
    \item \textbf{Synthases: }These enzymes catalyze the linking together of two molecules without the use of ATP, important in various biosynthetic pathways. These enzymes are labelled as Class \#4 in the dataset.
    \item \textbf{Ion channel receptors (ICRs):} Proteins that form pores in the cell membrane, allowing specific ions to pass through, contributing to a variety of cellular processes including the generation of electrical signals in neurons. These protein structures are labelled as Class \#5 in the dataset.  
    \item \textbf{Transcription factors (TFs):} Proteins that bind to specific DNA sequences and regulate the transcription of genetic information from DNA to messenger RNA. They often contain DNA-binding domains and activation domains that interact with other proteins. These  sequences are labelled as Class \#6 in the dataset. 
\end{enumerate}

Each of these families plays distinct roles in cellular function, and their structural differences are integral to their specific mechanisms of action. Since the detailed structural and bio-chemical behaviour of these proteins are out of the scope of this study, interested readers can refer to a biology textbook such as \citet{Alberts:2014}.

\begin{figure*}[!h]
\centering
\begin{minipage}{0.45\textwidth}
\includegraphics[width=\linewidth]{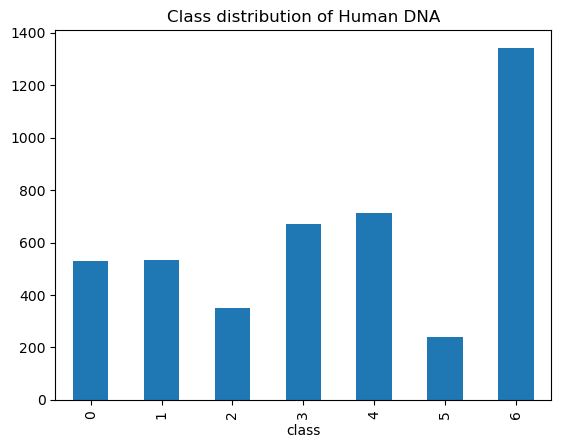}
\subcaption{Sequence Class distribution in the human genome.}
\label{fig:dist_human}
\end{minipage}
\hfill
\begin{minipage}{0.45\textwidth}
\includegraphics[width=\linewidth]{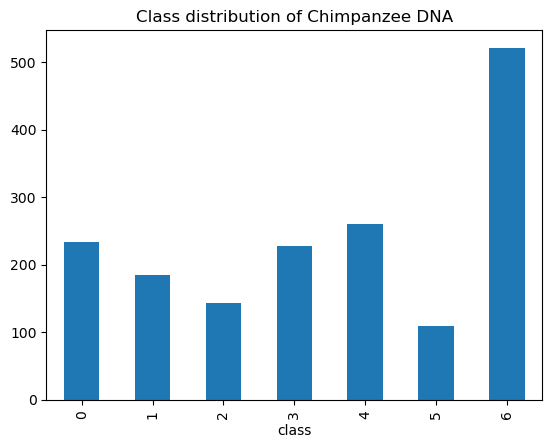}
\subcaption{Sequence Class distribution in the chimpanzee genome.}
\label{fig:dist_chimpanzee}
\end{minipage}
\\
\begin{minipage}{0.45\textwidth}
\includegraphics[width=\linewidth]{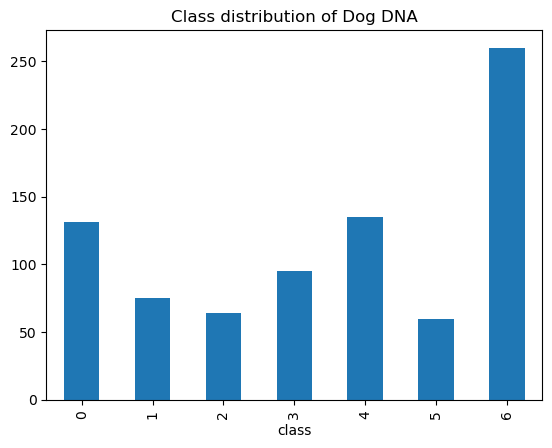}
\subcaption{Sequence Class distribution in the human genome.}
\label{fig:dist_dog}
\end{minipage}
\caption{Sequence Class distribution in the genomes of the species namely the human (\subref{fig:dist_human}), chimpanzee (\subref{fig:dist_chimpanzee}) and dog genomes (\subref{fig:dist_dog}).}
\label{fig:class_distribution_species}
\end{figure*}

\begin{figure}[!h]
\centering
\includegraphics[width=0.75\linewidth]{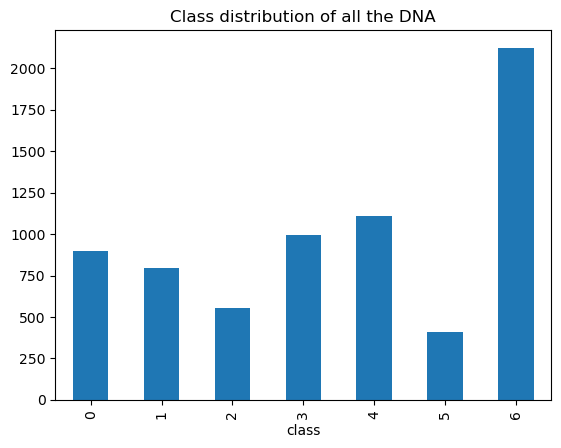}
\caption{Sequence Class distribution in the whole gene pool including the human, chimpanzee and dog genomes.}
\label{fig:dist_all}
\end{figure}

For the experiments we created a dataset by combining the given labelled DNA sequences. Total number of different gene families for each of three species can be seen in Table \ref{table:data-numbers} . By applying an 80\% - 20\% split scheme for each class after shuffling, a train and test datasets of size 5505 and 1377 are obtained respectively. The distribution over class labels and corresponding species can further be seen in Table \ref{table:gene-family-distribution} and Figure \ref{fig:class_distribution_species} respectively. Figure \ref{fig:dist_all} shows the total number of corresponding sequence classes in the whole gene pool consisting of all the information gathered from the human, chimpanzee and dog genomes.

\begin{table}[!h]
\caption{Dataset Composition by Species}
\centering
\begin{tabular}{|l|c|c|}
\hline
\textbf{Species} & \textbf{Train Size} & \textbf{Test Size} \\
\hline
Chimpanzee & 1345 & 337 \\
\hline
Human & 3504 & 876 \\
\hline
Dog & 656 & 164 \\
\hline
\textbf{Total} & 5505 & 1377 \\
\hline
\end{tabular}
\label{table:data-numbers}
\end{table}

\begin{table*}[!h]
\caption{Distribution of Gene Family Samples by Species}
\centering
\begin{tabular}{|l|c|c|c|c|c|}
\hline
\textbf{Gene Family} & \textbf{Class Label} & \textbf{Chimpanzee} & \textbf{Human} & \textbf{Dog} & \textbf{Total} \\
\hline
GPCRs & 0 & 234 & 531 & 131 & \textbf{896} \\
\hline
RTKs & 1 & 185 & 534 & 75 & \textbf{794} \\
\hline
PTPs & 2 & 144 & 349 & 64 & \textbf{557} \\
\hline
Synthetase & 3 & 228 & 672 & 95 & \textbf{995} \\
\hline
Synthase & 4 & 261 & 711 & 135 & \textbf{1107} \\
\hline
ICRs & 5 & 109 & 240 & 60 & \textbf{409} \\
\hline
TFs & 6 & 521 & 1343 & 260 & \textbf{2124} \\
\hline
\end{tabular}
\label{table:gene-family-distribution}
\end{table*}

\subsection{Classification Method with Compressors}

Compression algorithms can capture similarities between two texts by analyzing patterns and repetitions within the data. These algorithms often use techniques like dictionary-based encoding, where frequently occurring patterns are replaced with shorter representations. For instance, if two texts share common phrases or sequences, the algorithm will encode these similarities efficiently, using less space. This process inherently highlights the similarities between texts, as repeated patterns or sequences are compressed in a similar manner. The effectiveness of capturing these similarities depends on the algorithm's design, with some being more adept at identifying and utilizing these patterns for efficient compression.

\begin{figure*}[!h]
\centering
\resizebox{\textwidth}{!}{
\begin{tikzpicture}[node distance=2cm and 3cm, auto]
    \node [block] (comp1) {Compressor\\ $\mathtt{C}(x_1)$};
    \node [block, below of=comp1] (comp2) {Compressor\\ $\mathtt{C}(\mathtt{concat}(x_1,x_2))$};
    \node [block, below of=comp2] (comp3) {Compressor\\ $\mathtt{C}(x_2)$};
    \node [block, right=of comp2, fill=red!30] (ncd) {NCD\\(Equation \ref{eq:ncd})};
    \node [right=of ncd] (distance) {distance};
    \node [block, left=of comp2, fill=red!10] (concat) {concat};

    \draw [line] (comp1) -- (ncd);
    \draw [line] (comp3) -- (ncd);
    \draw [line] (comp2) -- (ncd);
    \draw [line] (concat) -- (comp2);
    \draw [line] (ncd) -- (distance);
    
    \node [left=of comp1] (x1) {$x_1$};
    \node [left=of comp3] (x2) {$x_2$};
    
    \draw [line] (x1) -- (comp1);
    \draw [line] (x1) -- (concat);
    \draw [line] (x2) -- (concat);
    \draw [line] (x2) -- (comp3);
    
    \draw [line] (comp1) -- (comp2);
    \draw [line] (comp3) -- (comp2);
    \draw [line] (comp2) -- (ncd);
\end{tikzpicture}
}
\caption{Flowchart illustrating the process of computing the Normalized Compression Distance (NCD) between two sequences $x_1$ and $x_2$ using a compressor and subsequent analysis.}
\label{fig:ncd_flowchart}
\end{figure*}
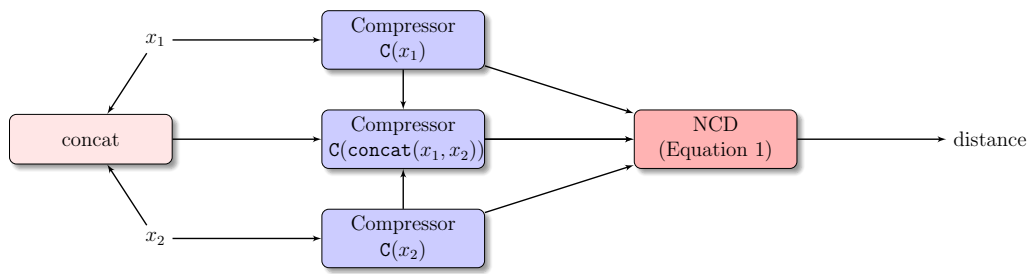

The classification method utilizes the normalized compression distance (NCD) \citet{Li:2004} given as Equation \ref{eq:ncd}. NCD is calculated by comparing the compression amounts of two strings and their concatenated versions. The flowchart of NCD can be depicted as Figure \ref{fig:ncd_flowchart}. 

\begin{equation}
\mathtt{NCD}(x_1, x_2) = \frac{\mathtt{C}(\mathtt{concat}(x_1,x_2)) - \min\{\mathtt{C}(x_1), \mathtt{C}(x_2)\}}{\max\{\mathtt{C}(x_1), \mathtt{C}(x_2)\}}\label{eq:ncd}
\end{equation}

\begin{table*}[!h]
\caption{Comparison of Compression Algorithms in DNA Sequence Classification in terms of computation time, accuracy, recall, precision and f1 score values.}
\centering
\resizebox{\textwidth}{!}{
\begin{tabular}{|l|c|c|c|c|c|c|}
\hline
\textbf{Algorithm} & \textbf{Computation Time (seconds)} & \textbf{Accuracy} & \textbf{Recall} & \textbf{Precision} & \textbf{F1 Score} \\
\hline
Gzip & 1735.81 & 0.962 & 0.962 & 0.963 & 0.962 \\
\hline
Snappy & {1726.49} & 0.932 & 0.932 & 0.933 & 0.932 \\
\hline
Brotli & 12551.60 & \textbf{0.966} & \textbf{0.966} & \textbf{0.967} & \textbf{0.966} \\
\hline
LZ4 & 1618.27 & 0.942 & 0.942 & 0.943 & 0.942 \\
\hline
Zstandard & \textbf{1560.72} & 0.930 & 0.930 & 0.935 & 0.931 \\
\hline
BZ2 & 2657.35 & 0.924 & 0.924 & 0.924 & 0.924 \\
\hline
LZMA & 9486.67 & 0.958 & 0.958 & 0.958 & 0.958 \\
\hline
\end{tabular}\label{table:compression-comparison}
}
\end{table*}

\begin{figure}[!h]
\centering
\includegraphics[width=0.75\linewidth]{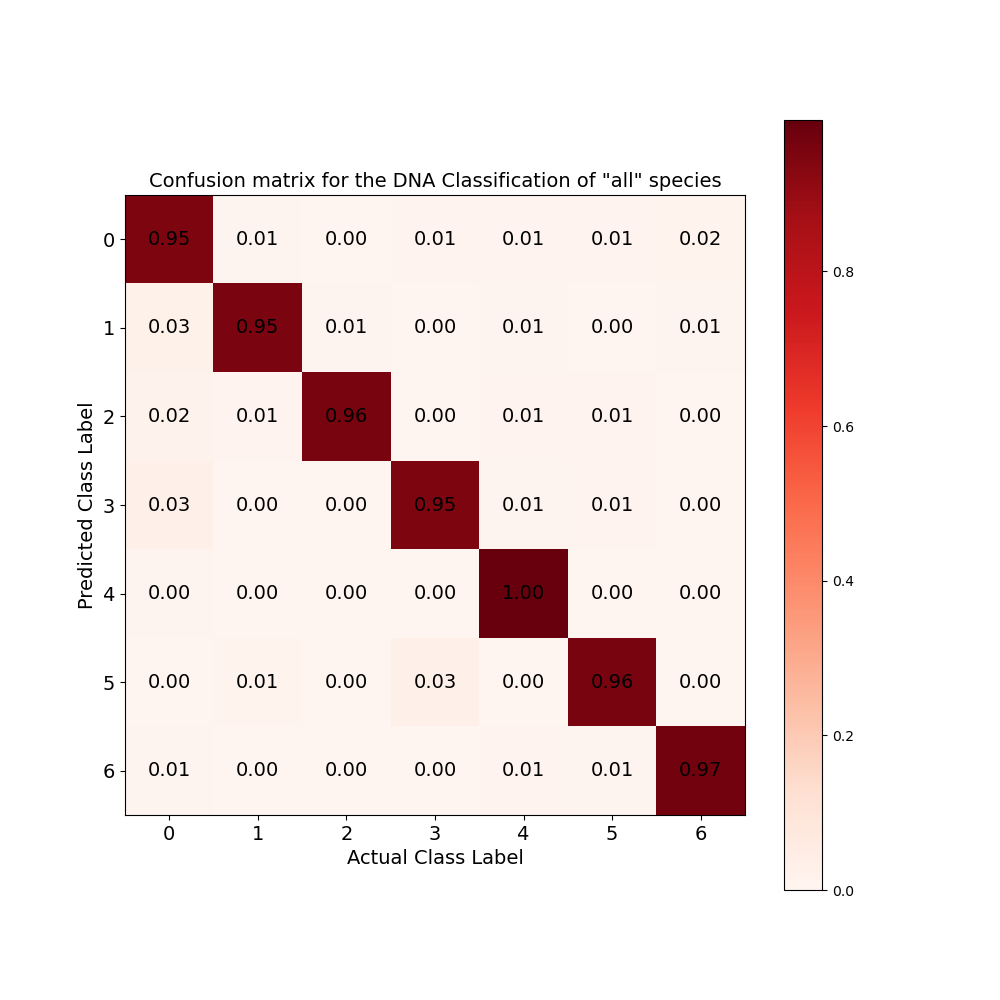}
\caption{Confusion Matrix for the Entire Dataset. This matrix provides an overview of the classifier's performance on the complete dataset, illustrating the accuracy of classification across all categories.}
\label{fig:confusion_whole_dataset}
\end{figure}

\begin{table*}[!t]
\caption{Performance Metrics of Compression Algorithms on DNA Sequence Classification for Different Species}
\centering
\resizebox{\textwidth}{!}{
\begin{tabular}{|c|c|c|c|c|c|c|c|c|c|c|c|c|}
\hline
\multirow{2}{*}{\textbf{Algorithm}} & \multicolumn{4}{c|}{\textbf{Chimpanzee}} & \multicolumn{4}{c|}{\textbf{Human}} & \multicolumn{4}{c|}{\textbf{Dog}} \\
\cline{2-13}
& \textbf{Accuracy} & \textbf{Precision} & \textbf{Recall} & \textbf{F1 Score} & \textbf{Accuracy} & \textbf{Precision} & \textbf{Recall} & \textbf{F1 Score} & \textbf{Accuracy} & \textbf{Precision} & \textbf{Recall} & \textbf{F1 Score} \\
\hline
Gzip & 0.997 & 0.997 & 0.997 & 0.997 & 0.944 & 0.945 & 0.944 & 0.944 & 0.988 & 0.989 & 0.988 & 0.988 \\
\hline
Snappy & 0.994 & 0.994 & 0.994 & 0.994 & 0.943 & 0.944 & 0.943 & 0.943 & 0.744 & 0.748 & 0.744 & 0.745 \\
\hline
Brotli & \textbf{1.000} & \textbf{1.000} & \textbf{1.000 }& \textbf{1.000 }& \textbf{0.947} & \textbf{0.950} & \textbf{0.947} & \textbf{0.948} & \textbf{0.994 }& \textbf{0.994} & \textbf{0.994} & \textbf{0.994} \\
\hline
LZ4 & 0.997 & 0.997 & 0.997 & 0.997 & 0.944 & 0.946 & 0.944 & 0.944 & 0.817 & 0.829 & 0.817 & 0.819 \\
\hline
Zstandard & 0.994 & 0.994 & 0.994 & 0.994 & 0.900 & 0.909 & 0.900 & 0.901 & 0.963 & 0.967 & 0.963 & 0.963 \\
\hline
BZ2 & 0.985 & 0.985 & 0.985 & 0.985 & 0.893 & 0.894 & 0.893 & 0.893 & 0.963 & 0.966 & 0.963 & 0.964 \\
\hline
LZMA & 1.000 & 1.000 & 1.000 & 1.000 & 0.936 & 0.937 & 0.936 & 0.936 & 0.988 & 0.989 & 0.988 & 0.988 \\
\hline
\end{tabular}
}
\label{table:wide-compression-performance}
\end{table*}

\begin{figure*}[!h]
\centering
\begin{minipage}{0.45\textwidth}
\includegraphics[width=\linewidth]{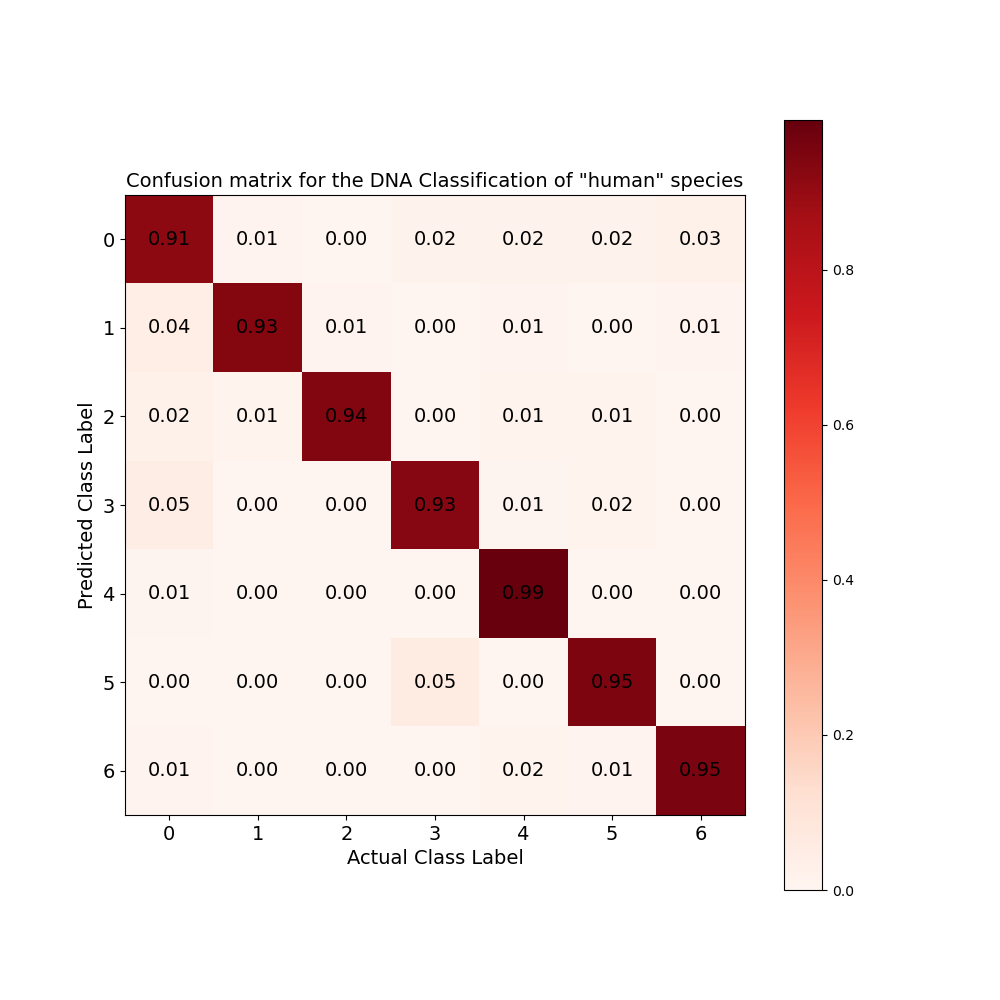}
\subcaption{Confusion matrix for human DNA sequences}
\label{fig:confusion_human}
\end{minipage}
\hfill
\begin{minipage}{0.45\textwidth}
\includegraphics[width=\linewidth]{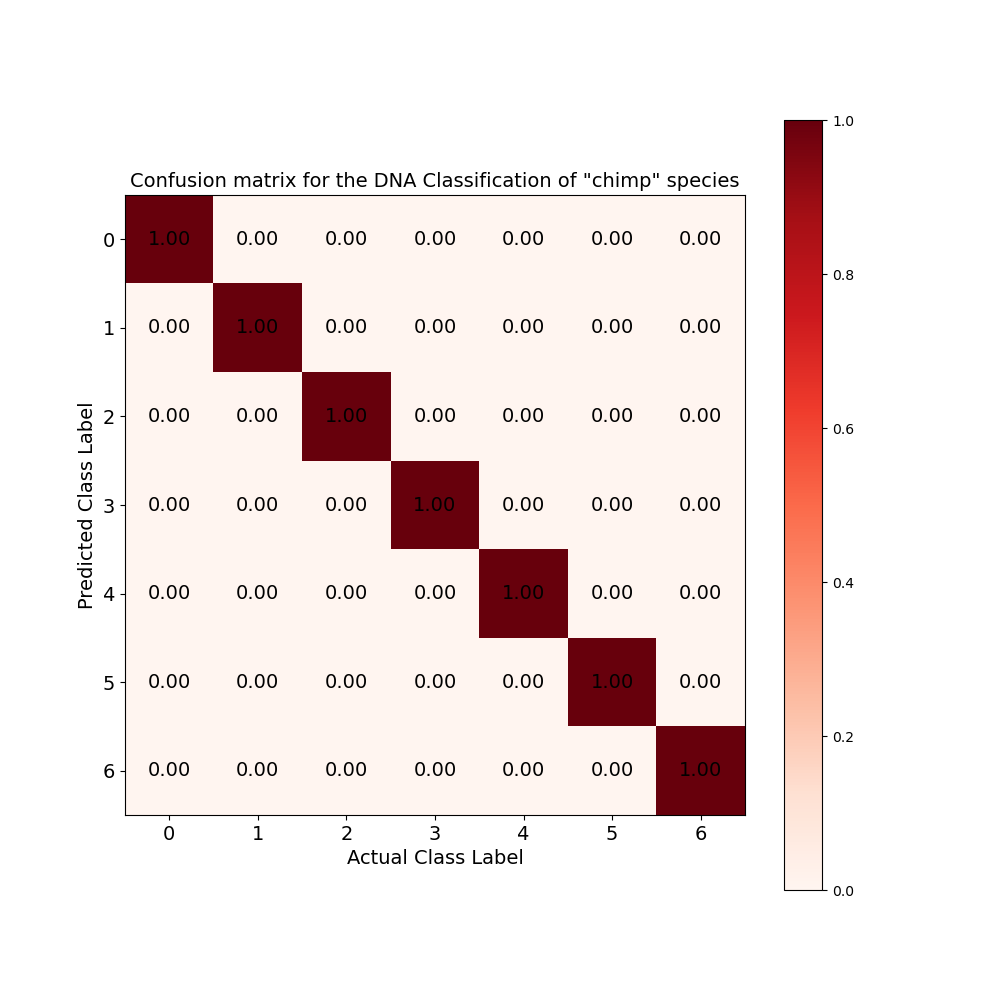}
\subcaption{Confusion matrix for chimpanzee DNA sequences}
\label{fig:confusion_chimpanzee}
\end{minipage}
\\
\begin{minipage}{0.45\textwidth}
\includegraphics[width=\linewidth]{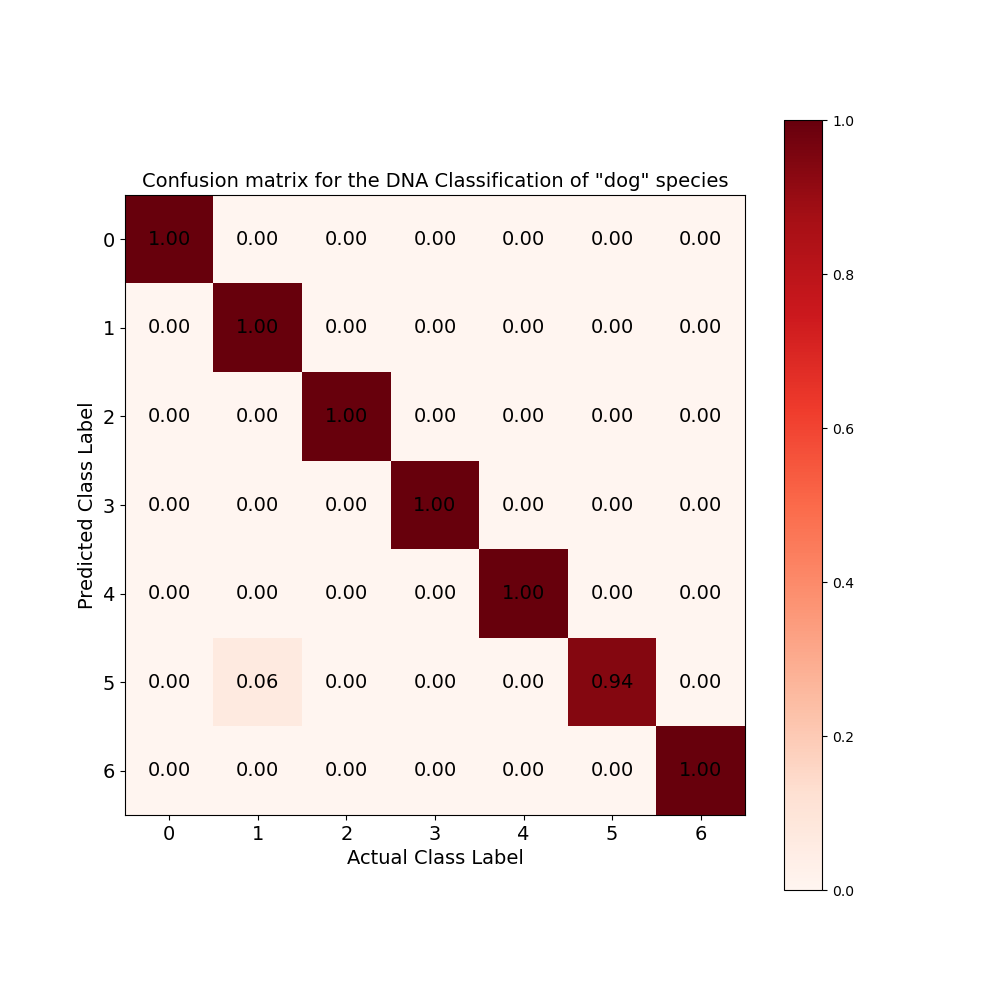}
\subcaption{Confusion matrix for dog DNA sequences}
\label{fig:confusion_dog}
\end{minipage}
\caption{Confusion matrices for subspecies-wise classifications: Human, Chimpanzee, and Dog DNA. These matrices detail the classifier's accuracy for each subspecies, highlighting the precision in distinguishing between the seven genomic classes labelled as `Class 0' to `Class 6'.}
\label{fig:confusion_subspecies}
\end{figure*}

\section{Experiments}
\subsection{Compression algorithms}

In this work, we aim to prove the applicability of the NCD based classification algorithm \citet{Jiang:2023} to DNA sequence classification problem. Whilst doing that, we also try different compression algorithms which are conventionally used by simply importing corresponding libraries in  a Python script. Our selected compression algorithms are:

\begin{enumerate}
    \item \textbf{Gzip:} Gzip uses the DEFLATE algorithm, which is a combination of LZ77 \citet{Ziv:1977} and Huffman coding \citet{Huffman:1952}. It offers a good balance between compression ratio and speed, making it widely used for file compression and in web contexts.
  
    \item \textbf{Snappy:} Developed by Google, Snappy is based on LZ77. It does not aim for maximum compression or compatibility with any other compression library; instead, it aims for very high speeds and reasonable compression.Often used in systems like databases and interprocess communication where speed is more crucial than the degree of compression.

    \item \textbf{Brotli:} Brotli is a mix of a modern variant of the LZ77 algorithm, Huffman coding, and 2nd order context modeling. It provides higher compression ratios than Gzip, especially for text data. It's somewhat slower in compression but still very fast in decompression. Commonly used in web traffic for assets like HTML, CSS, and JavaScript due to its efficient compression of text.

    \item \textbf{LZ4:} LZ4 is based on LZ77 but is designed for very fast compression and decompression. It trades off compression ratio for speed, being one of the fastest compressors in terms of both compression and decompression. Ideal for scenarios where processing speed is more important than reducing data size, such as real-time applications.

    \item \textbf{Zstandard (Zstd):} Zstd also combines LZ77 with Huffman coding but introduces new techniques like a fast dictionary-based compression.Offers a wide range of compression levels, providing a balance between speed and compression ratio. It's adaptable to different types of data.Versatile for various scenarios, from real-time communication to archival storage, due to its scalability in compression levels.

    \item \textbf{BZ2 (Bzip2):} Bzip2 uses the Burrows-Wheeler transform \citet{Burrows:1994} followed by the Move-To-Front transform \citet{Bentley:1986} and Huffman coding. Generally provides higher compression ratios than Gzip, but at the cost of slower compression speed. Suitable for applications where compression ratio is more important than compression or decompression speed.
    
    \item \textbf{LZMA (Lempel-Ziv-Markov chain algorithm):} LZMA uses a dictionary compression scheme (like LZ77) and a range coding (similar to arithmetic coding). Known for its very high compression ratios, but with slower compression and decompression speeds compared to others like Gzip or LZ4. Ideal for applications where the compressed size is the primary concern, and time is not a constraint, such as compressing large backups or software distribution.
\end{enumerate}

Each of these algorithms has its unique strengths and weaknesses. The choice of which one to use depends on the specific requirements of the application, such as the need for fast compression, fast decompression, or maximizing the compression ratio. For the specific problem of this work we also validate and show the trade-off between computation speed and accuracy in Section \ref{sec:results}.

\subsection{Computation}

To utilize the parallel processing capability of the computer, we used the 'pathos' python library. In the code segment shown in Listing \ref{listing:python}, two main functions to construct the distance matrix can be seen. 

$\mathtt{create\_distance\_matrix}$ utilizes pooling and further accepts a variable for selecting the compression algorithm. $\mathtt{distance\_calc}$ function is used to calculate the normalized compression distance (see Equation \ref{eq:ncd}) for a given pair of training and test sample.

The experiments are performed on a computer with M1 chip-set and 8 cores. A Python notebook and a Conda environment were used for simplicity, portability and reproducibility. The notebook and the Conda environment configuration files are shared in our repository \citet{Ozan:2023}.

Running the code as separate Python scripts is faster than executing the same scripts within Python notebooks. Nevertheless, the experimental setup provides insights into the running times of different compression algorithms. In Table \ref{table:compression-comparison} the computation times for each selected compression algorithms can be seen. 

\begin{figure*}[!h]
\begin{lstlisting}[language=Python, basicstyle=\ttfamily\tiny,caption=Python functions to construct the distance matrix. Elements of the matrix corresponds to a normalized compression distance between the samples from the training set and the test set., label=listing:python ]
def distance_calc(compression, test, train, i, j):
    test_string = test.iloc[i][`sequence']
    train_string = train.iloc[j][`sequence']
    if compression in [`gzip', `snappy', `brotli', `bz2']:
        compressor = globals().get(compression)
        Cx1 = len(compressor.compress(test_string))
        Cx2 = len(compressor.compress(train_string))
        Cx1x2 = len(compressor.compress((" ".join([test_string, train_string]))))
    elif compression == `lz4':
        Cx1 = len(lz4.frame.compress(test_string))
        Cx2 = len(lz4.frame.compress(train_string))
        Cx1x2 = len(lz4.frame.compress((" ".join([test_string, train_string]))))
    elif compression == `zstandard':
        Cx1 = len(zstd_compress(test_string.encode()))
        Cx2 = len(zstd_compress(train_string.encode()))
        Cx1x2 = len(zstd_compress((" ".join([test_string, train_string]))))
    elif compression == `lzma':
        Cx1 = len(lzma.compress(test_string.encode()))
        Cx2 = len(lzma.compress(train_string.encode()))
        Cx1x2 = len(lzma.compress((" ".join([test_string, train_string])))) 
    distance = (Cx1x2 - min(Cx1, Cx2)) / max(Cx1, Cx2)
    return [i, j, distance]
    
def create_distance_matrix(train, test, pool, compression=`gzip'):
    distance_matrix = np.zeros((len(test), len(train)))
    for i in range(len(test)):
        result = pool.map(partial(distance_calc, compression, test, train, i), range(len(train)))
        for j in result:
            distance_matrix[j[0]][j[1]] = j[2]
    return distance_matrix    
\end{lstlisting}
\end{figure*}

\section{Results}\label{sec:results}

The classification method in \citet{Jiang:2023} utilizes both NCD and K-nearest neighborhood classification. The number of neighbors is typically selected as 3. However, in our work, we opted for the single nearest neighbor, using its class information for prediction. The method yields results comparable to state-of-the-art techniques for the DNA sequence classification problem. Table \ref{table:compression-comparison} summarizes the experiment results. The confusion matrix of the proposed classifier algorithm utilizing Brotli compression for the whole dataset can be seen in Figure \ref{fig:confusion_whole_dataset}.

Regarding computation speed, the Zstandard' compression algorithm appears to be the fastest. Conversely, the `Brotli' compression, while being the slowest, demonstrates the best overall accuracy over the dataset with species-wise and gene-class-wise distribution given in Table \ref{table:gene-family-distribution}.

A more detailed accuracy comparison with species breakdown can be seen in Table \ref{table:wide-compression-performance}. We see that Brotli compression has the best performance over the three species' DNA samples. The depiction of the confusion matrices of three species, human, chimpanzee and dog, can be seen in Figures \ref{fig:confusion_human}, \ref{fig:confusion_chimpanzee} and \ref{fig:confusion_dog} respectively.

\section{Discussion and Conclusion}

This study addresses a bioinformatics problem related to DNA sequence classification. The nature of the problem typically restricts the dataset size, hence the problem is mostly considered not well fitted for deep learning. Researchers tend to prefer to use machine learning and NLP inspired methods such as k-mers. As a state-of-the-art result in \citet{Juneja:2022} the authors reported an overall accuracy of 98.4\% over the same data base we use in this study. Our NCD approach yielded a maximum overall accuracy of 96.6\% which makes the approach considerable for the DNA sequence classification problem.

\begin{figure*}[!h]
\centering
\begin{minipage}{0.49\textwidth}
\includegraphics[width=\linewidth]{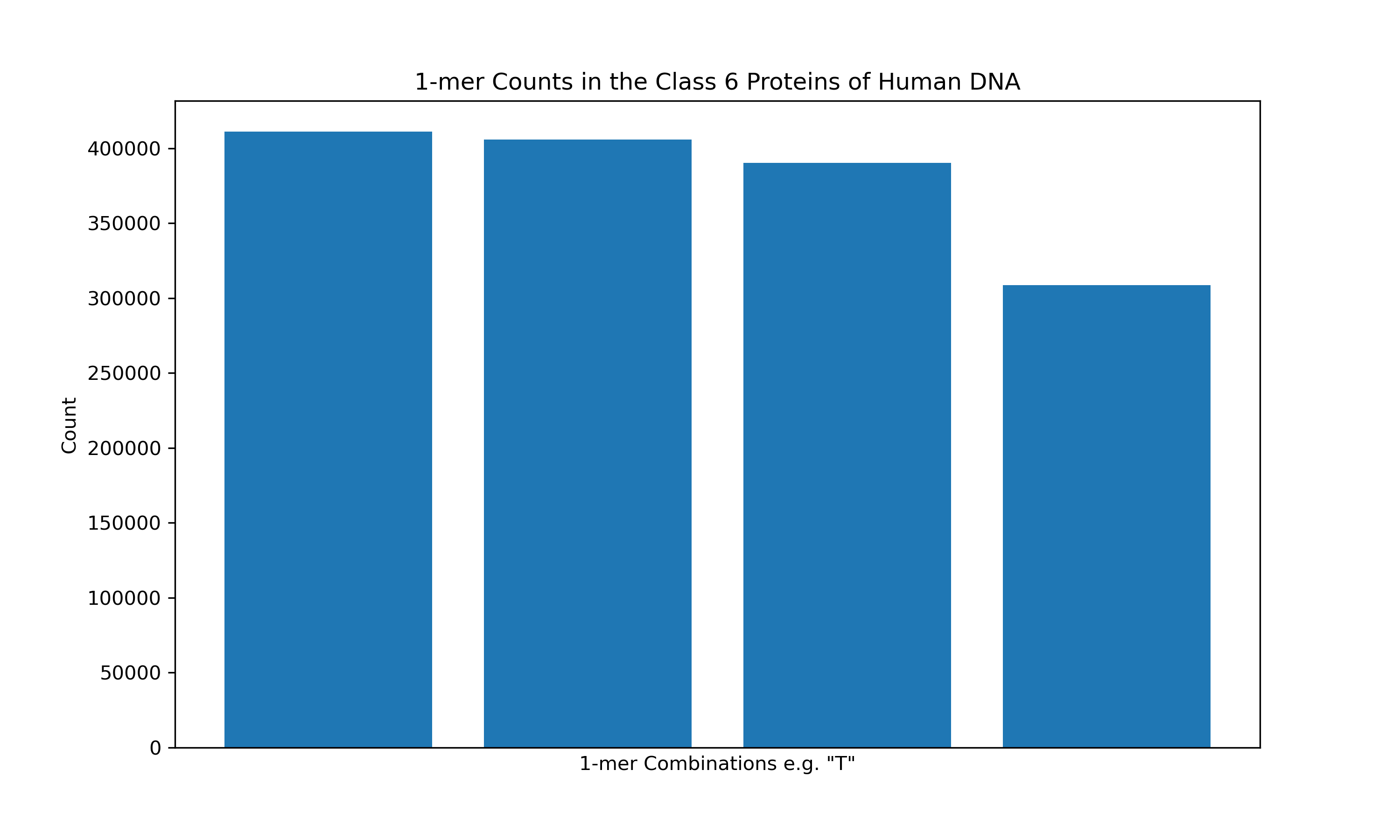}
\subcaption{Histogram of basic nucleotide bases in Human DNA sequences.}
\label{fig:class_6_1_mer}
\end{minipage}
\begin{minipage}{0.49\textwidth}
\includegraphics[width=\linewidth]{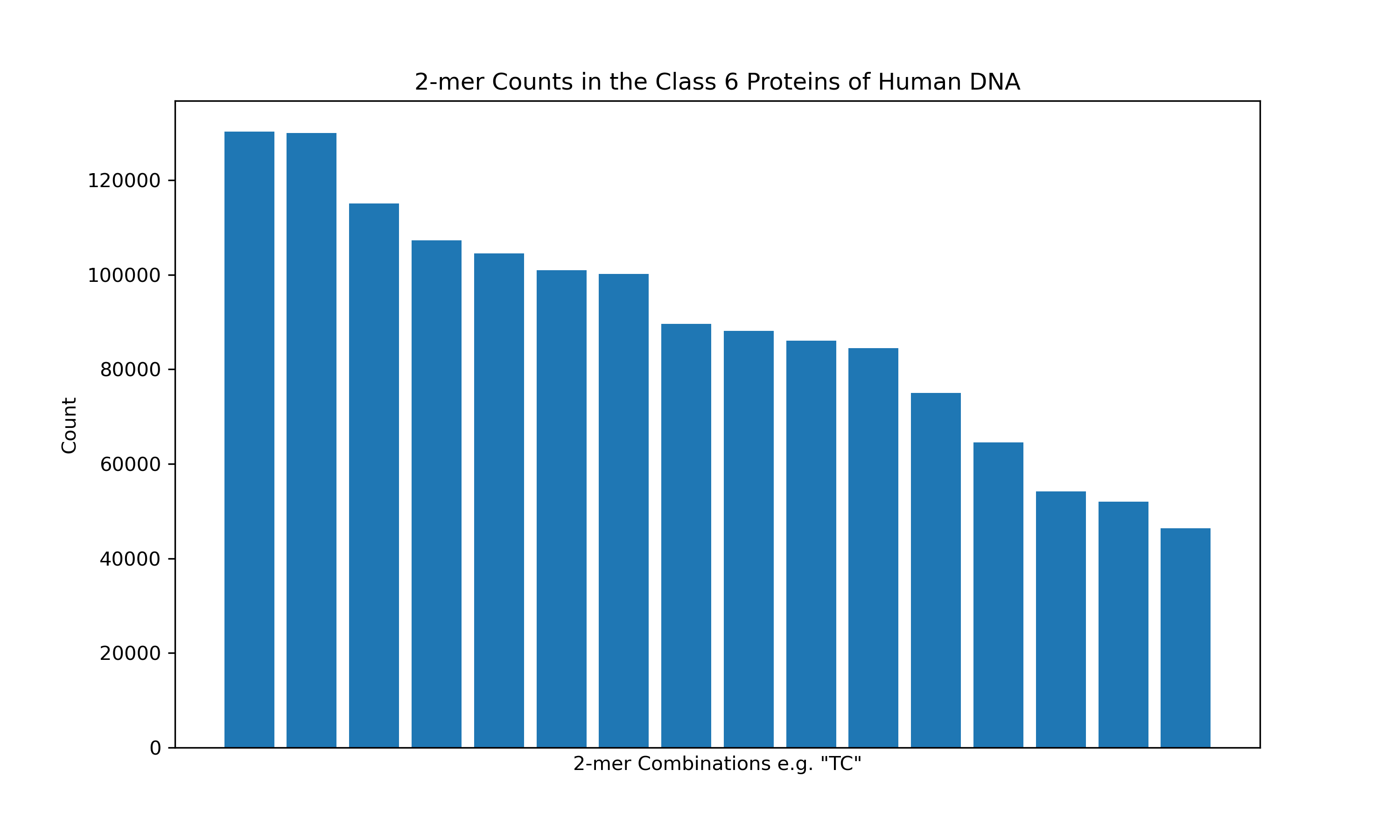}
\subcaption{Histogram of 2-mer nucleotide squences in Human DNA sequences.}
\label{fig:class_6_2_mer}
\end{minipage}
\\
\begin{minipage}{0.49\textwidth}
\includegraphics[width=\linewidth]{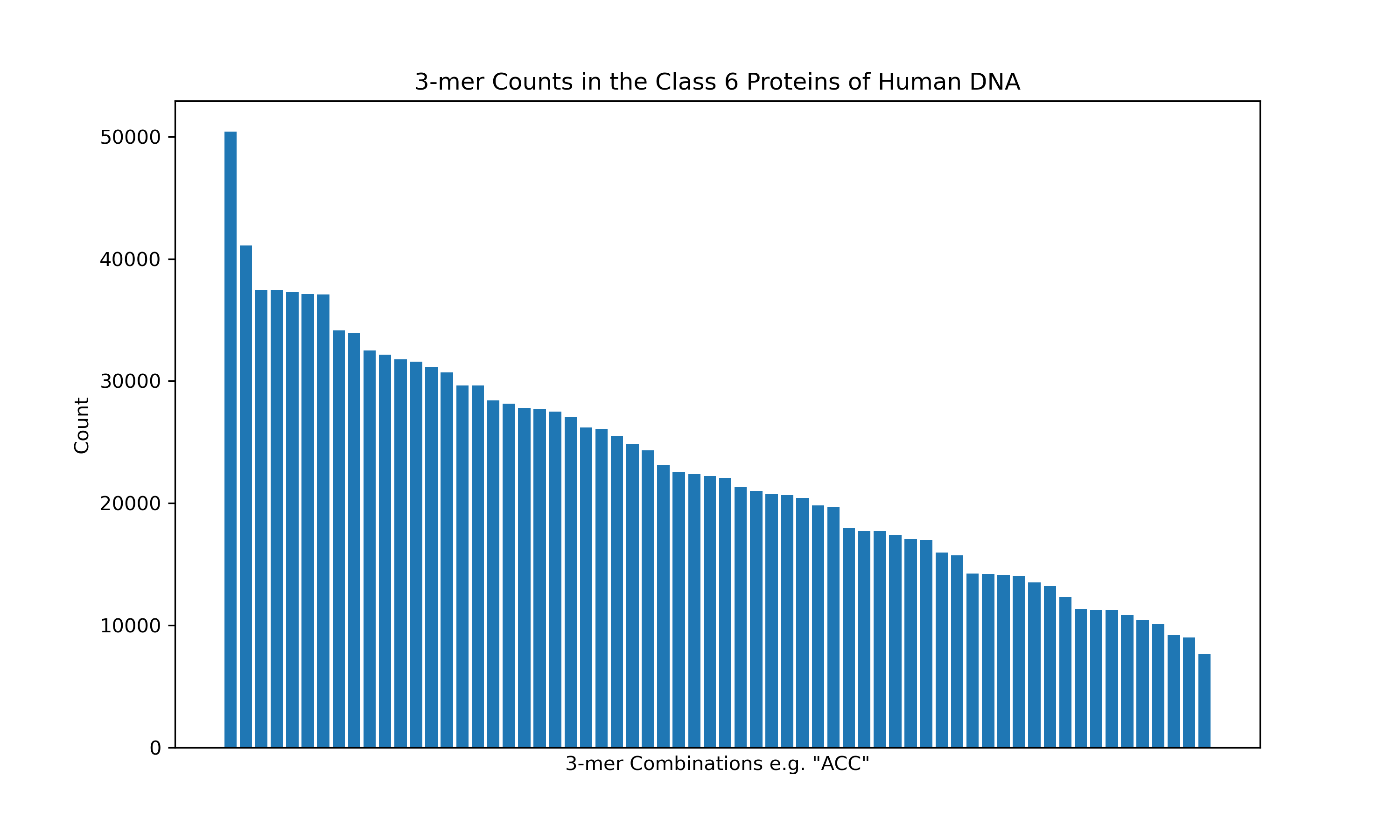}
\subcaption{Histogram of 3-mer nucleotide squences in Human DNA sequences.}
\label{fig:class_6_3_mer}
\end{minipage}
\begin{minipage}{0.49\textwidth}
\includegraphics[width=\linewidth]{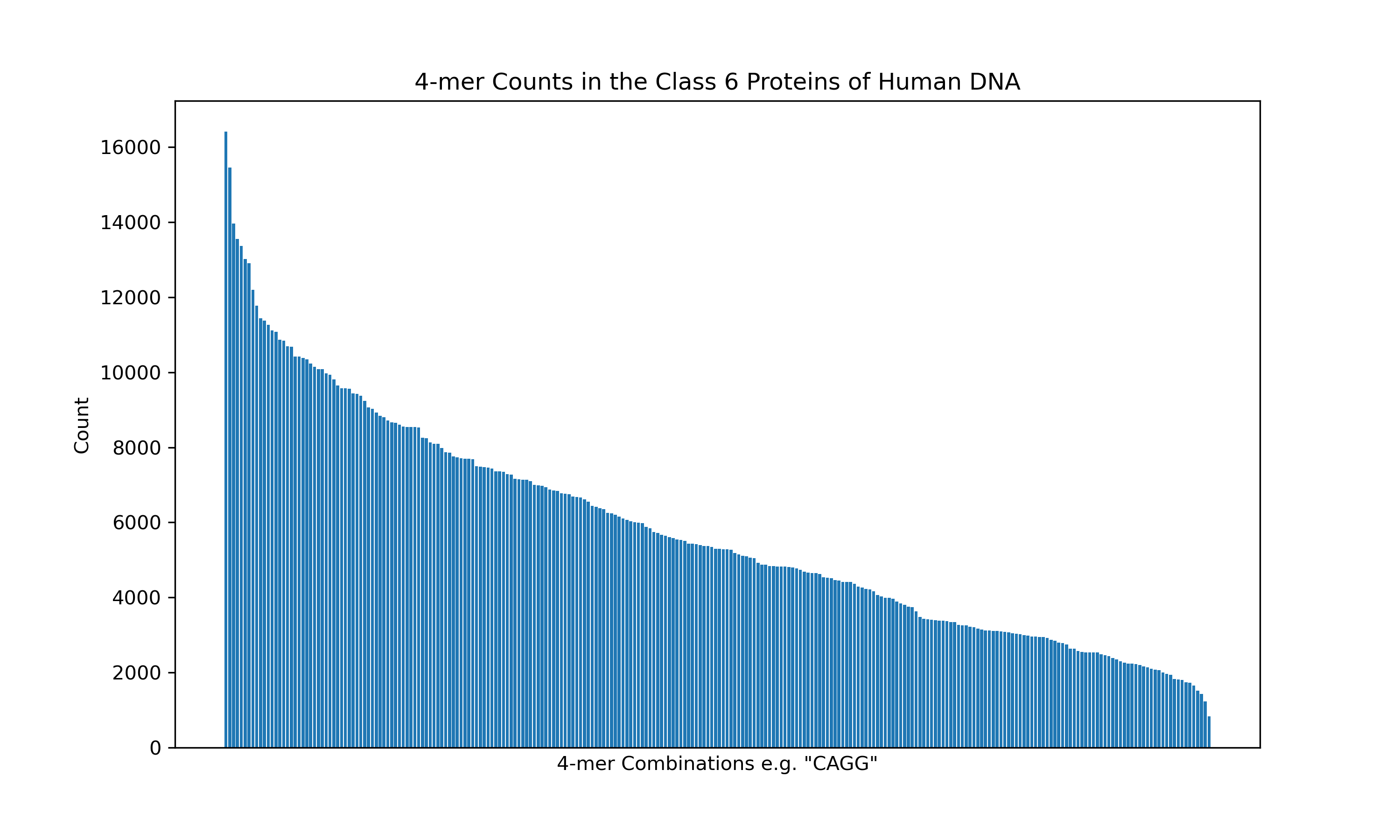}
\subcaption{Histogram of 4-mer nucleotide squences in Human DNA sequences.}
\label{fig:class_6_4_mer}
\end{minipage}
\caption{\subref{fig:class_6_1_mer} to \subref{fig:class_6_4_mer}: Shows the histograms of 1, 2, 3, and 4-mer nucleotide sequences in Human DNA sequences.}
\label{fig:human_class_6_n_mer_counts}
\end{figure*}

In Figure \ref{fig:human_class_6_n_mer_counts}, we can see the histogram of 1, 2, 3 and 4-mer nucleotide sequence histograms in Class 6 sequences of Human Genome (see Table \ref{table:gene-family-distribution}). From these figures it can be interpreted that as $n$ increases n-mers constitute more discriminative features. This is the reason why the n-mer based classification methods such as \citet{Juneja:2022} works successfully. These distributions also validate why the compression based methods work as well. The compressors are inherently robust in finding common patterns and compress the data accordingly. If there are more common segments in two given sample sequences, they compress more with respect to two sequences which share less common segments.

We saw that most of the chosen compression algorithms give over 90\% accuracy. However, the Brotli compressor achieves the highest accuracies across all species. However, despite its fast computation time, the Zstandard algorithm only reaches 89.3\% accuracy for human DNA and another fast compressor Snappy only reaches 74.4\% accuracy for dog DNA.

As highlighted in \citet{Jiang:2023}, the NCD approach's computational complexity is $O(n^2)$ which makes the method not suitable for very large datasets. As a further study, addressing the computational complexity of the NCD method and finding a method to make it also applicable for larger datasets can be an interesting work. As an alternative text classification method the NCD approach seems to be a good alternative also for DNA sequence classification problem of the bioinformatics literature.

\section*{Declaration of generative AI and AI-assisted technologies in the writing process}

During the preparation of this work the author used OpenAI ChatGPT in order to improve language and readability. After using this tool/service, the author reviewed and edited the content as needed and take full responsibility for the content of the publication.

\section*{Supplementary Material}

For the purpose of ensuring transparency and reproducibility in this research, the complete codebase and dataset used in this study are made publicly available. Interested researchers and practitioners can access these resources through the corresponding GitHub repository \citet{Ozan:2023}. The repository contains all data and source code used for the implementation of the proposed DNA sequence classification algorithm. This includes scripts for data preprocessing, model training, evaluation, and any additional utility functions. Accompanying the code, the dataset employed in this study is also available for download. This dataset is crucial for researchers looking to replicate this study or build upon it. Comprehensive documentation is provided to guide users through the setup, execution, and understanding of the code and data.  We welcome contributions and collaborations from the community. Researchers can utilize the repository as a starting point for their investigations or to compare with alternative methodologies in DNA sequence classification or other similar problems.

\bibliographystyle{plainnat}
\bibliography{refs}
\end{document}